# The microscopic origin of the spin-induced linear and quadratic magnetoelectric effects

Maocai Pi[1,2], Xifan Xu[1,2], Mingquan He[1,2], Yisheng Chai[1,2*]

[1]*Center of Quantum Materials and Devices, Chongqing University, Chongqing 401331, China.*

[2]*Low Temperature Physics Laboratory, College of Physics, Chongqing University, Chongqing 401331, China*

*Corresponding emails: yschai@cqu.edu.cn

Understanding of the intimate cross-coupling between electric and magnetic degrees of freedom in solids usually requires sophisticated models and time-consuming calculation methods. Instead of macroscopic symmetry analysis, we present a simple but general approach to explore the microscopic mechanism of magnetoelectric (ME) effects in magnetic ordered materials based on local spin and lattice symmetry analysis. Our methods are successfully applied to $Cr_2O_3$ and orthorhombic $R$MnO$_3$ with linear and quadratic ME effects, respectively. We revealed all the possible microscopic origins of every non-zero ME coefficients that cannot be easily fulfilled by other theoretical methods. Moreover, the contribution from each mechanism can be located down to specific spins or spin pairs. Our theoretical approach is capable of providing a detailed guide to explore rich spin-induced magnetoelectrics with strong ME effects.

**Introduction**

Crystals with long-range magnetic orders can lead to emergent phenomena as cross-coupling between electric and magnetic degrees of freedom, e.g., magnetoelectric (ME) effect or spin-induced polarization (type-II multiferroics). Based on the global symmetry of magnetic ordering, a macroscopic polarization (*P*) can be induced if the space inversion symmetry is broken. A linear ME effect is allowed if both time reversal and space inversion symmetries are broken but their combinations are preserved. So far, the microscopic origins of spin-induced polarization have been mainly studied by First-principles calculation, and are attributed to three main mechanisms, including two-spin mechanisms (exchange-striction, inverse Dzyaloshinskii–Moriya (DM) coupling mechanism) and single-spin mechanism (*p-d* hybridization mechanism) [1-3]. However, the local symmetric information in type-II multiferroics have not been fully utilized. Nevertheless, a lattice site-symmetry approach was introduced to find the microscopic mechanisms in high crystal symmetric systems [4-7]. For example, $LaMn_3Cr_4O_{12}$ with antiferromagnetic ordering in cubic lattice is resolved to be antisymmetric exchange-striction mechanism [8].

The ME effects can be linear (LME) or quadratic (QME), *i.e.*, the electric polarization *P* (magnetization *M*) responses linearly or quadratically to the application of magnetic field *H* (electric field *E*). The corresponding coupling coefficients can be defined as $\alpha = P/H$ ($\mu_0 M/E$) or $\beta = P/H^2$ ($\mu_0 M/E^2$), respectively. As a new physical principle to realize novel electronic devices, the study of the ME effects has attracted considerable interests [9,10]. Despite of remarkable progresses in composite ME materials and related devices [11], the ME devices made by spin-induced ME materials are still very rare due to their weak coupling strength [12]. In addition, the microscopic origin of ME effects in the magnetic-ordered systems are also not well studied. The most typical LME material $Cr_2O_3$ in antiferromagnetic (AFM) order has a maximum $\alpha$ value of 7 ps/m [13], while the orthorhombic $R$MnO$_3$ ($R$ is rare earth ions) with E-type AFM ground states shows a quadratic response of $P$ to $H$ [14,15] with a small $\beta$ value. Besides the $Cr_2O_3$, the microscopic mechanisms of linear or quadratic ME effects are less explored in other single-phase materials due to the lack of theoretical treatments. To find more magnetic systems with giant ME effects, a universal theoretical approach is highly desired to reveal the microscopic origins of LME and QME effects or even higher order effects.

In this letter, we developed a general theoretical approach to understand the microscopic origin of LME and QME effects simultaneously. Based on the site-symmetries of lattice and spin together, we elucidate the role of each mechanism and each spin pairs in the LME effects in $Cr_2O_3$ which is omitted previously. In orthorhombic $R$MnO$_3$, we reveal that its QME coupling coefficients in E-type AFM state have two dominating non-zero components governed only by exchange striction mechanism. Our comprehensive approach could also be applicable to other single phase ME materials and to the ME effects above the second order.

From atomic scale, the microscopic origin of electric dipoles, whether they are induced from the type-II multiferroics or from the magnetic ordered magnetoelectrics under $H$, would not have much difference. To reveal such similarity qualitatively, one can always describe the local electric dipole $p = p_{12}+p_{11}+p_{22}$ from a spin-pair $S_1$ and $S_2$ irrespective of multiferroics or magnetoelectrics. In general, the coupling between $S_1$ and $S_2$ produces the local electric-dipole $p_{12}$, while the local-dipoles $p_{11}$ and $p_{22}$ are produced by spins $S_1$ and $S_2$, respectively, with its surrounding ligand atoms. Both $p_{ii}$ and $p_{ij}$ ($i, j = 1,2$) can be expressed as functions with the two variables $S_1$ and $S_2$ under the second order Taylor expansion in the Einstein convention:

$$p_{ii}^\gamma = P_{ii}^{\alpha\beta\gamma} S_i^\alpha S_i^\beta, \quad p_{ij}^\gamma = P_{ij}^{\alpha\beta\gamma} S_i^\alpha S_j^\beta, \qquad (1)$$

where superscripts $\alpha$, $\beta$ and $\gamma$ represent the Cartesian coordinates $x$, $y$, $z$, $P_{ii}^{\alpha\beta\gamma}$ and $P_{ij}^{\alpha\beta\gamma}$ are the local third-rank magnetoelectric tensors corresponding to spin-pair and single spin with ligand atoms, respectively. Only even-order terms in the coefficients of Taylor expansion are reserved because of time reversal symmetry. The macroscopic $P = (P^x, P^y, P^z)$ is:

$$P^\gamma = \sum_{i,j} P_{ij}^{\alpha\beta\gamma} S_i^\alpha S_j^\beta, \qquad (2)$$

where $i, j$ run over all spins and adjacent spin pairs. This local ME tensor analysis has been successfully applied to explain the microscopic origin of $P$ in multiferroics in the absence of magnetic field $H$ [4-8] (see Supplemental Material (SM) for more details [16]). When a small magnetic field is applied, the change of spin configuration in either type-II multiferroics or magnetoelectrics would induce a small change in polarization $\delta P = P(H)-P(0\ T)$, which is indeed the ME effect.

The spin vector $S_i(H)$ at atom $i$ can be expanded as, in the Einstein convention:

$$S_i(H) = S_i(0) + \frac{\partial S_i^\alpha}{\partial H^l} H^l + \frac{\partial^2 S_i^\alpha}{2\partial H^l \partial H^m} H^l H^m + \cdots, \left(\frac{\partial S_i^\alpha}{\partial H^l} = A_i^{\alpha l}, \frac{\partial^2 S_i^\alpha}{2\partial H^l \partial H^m} = B_i^{\alpha lm}\right), \qquad (3)$$

where $\alpha$, $l$ and $m$ run over the Cartesian coordinates $x$, $y$, $z$. $A_i^{\alpha l}$ and $B_i^{\alpha lm}$ are linear and quadratic local spin susceptibility tensors under small $H$, respectively. Based on the symmetry analysis, $A_i^{\alpha l}$ is invariant under both space inversion and time reversal symmetry, and has the same transformation properties as the macroscopic susceptibility tensor. On the other hand, $B_i^{\alpha lm}$ is invariant with space inversion symmetry but opposite with time reversal symmetry, which has the same transformation properties as the macroscopic piezomagnetic coefficient tensor.

The above spin variations lower the symmetry of the magnetic materials, which causes the variation of $\delta P$ by putting Eq. (3) into Eq. (2):

$$P^\gamma(H) = \sum_{i,j} P_{ij}^{\alpha\beta\gamma}[(S_i^\alpha + A_i^{\alpha l}H^l + B_i^{\alpha lm}H^lH^m + \cdots)(S_j^\beta + A_j^{\beta l}H^l + B_j^{\beta lm}H^lH^m + \cdots)]$$

$$= \sum_{i,j} P_{ij}^{\alpha\beta\gamma} S_i^\alpha S_j^\beta + \sum_{i,j} P_{ij}^{\alpha\beta\gamma}\left(S_j^\beta A_i^{\alpha l} + S_i^\alpha A_j^{\beta l}\right)H^l + \sum_{i,j} P_{ij}^{\alpha\beta\gamma}\left(S_i^\alpha B_j^{\beta lm} + S_j^\beta B_i^{\alpha lm} + A_i^{\alpha l}A_j^{\beta m}\right)H^lH^m + \cdots \quad (4)$$

The first term is polarization without $H$, which is zero for pure ME systems like $Cr_2O_3$ but is finite for multiferroics like o-$R$MnO$_3$ with E-type AFM. The second and third terms describe the $\delta P$ induced by LME and QME effects, respectively. Then, the mathematical expressions of related ME tensor can be derived from Eq. (4) accordingly. All the terms and related information to calculate each macroscopic property are summarized in Table I. Furthermore, we are able to infer the microscopic mechanisms of ME effects from the nonzero components of $P_{ii}^{\alpha\beta\gamma}$ and $P_{ij}^{\alpha\beta\gamma}$ in $\delta P$ (see SM for details [16]).

We first apply this approach to the most well-known LME material, $Cr_2O_3$. The AFM order appears below $T_N$=307 K, showing in FIG. 1(a). Its magnetic point group is **-3'm'** with the symmetry operations **m'** $\perp x$, **-3'**//$z$. According to the Neumann's principle, the non-zero LME coefficients $\alpha^{ij}$ are $\alpha^{xx}=\alpha^{yy}=\alpha_\perp$ and $\alpha^{zz}=\alpha_\parallel$ whose variations strongly rely on the magnetic susceptibilities. The temperature dependent $\alpha_\perp$ decreases smoothly upon cooling, whereas $\alpha_\parallel$ shows a peak below $T_N$ which drops quickly and becomes negative as approaching to 0 K which comes from Van-fleck paramagnetism [13,17]. The positive peak feature below $T_N$ is considered to be mainly due to the exchange striction mechanism [18]. The agreement between other existing theories and experiments is very well in the case of $\alpha_\perp$ but fails in that of $\alpha_\parallel$ [19]. These calculations, however, cannot tell all the possible microscopic origins and detailed locations responsible for the observed LME effects in this system [18].

To resolve the problems mentioned above, we evaluate the LME components $\alpha_\perp$ and $\alpha_\parallel$ from our theory. The starting AFM at $H=0$ would be: $S_1=S_4=-S_2=-S_3=(0,0,S^z)$, where the magnetic moments of Cr spins $S_1$, $S_2$, $S_3$ and $S_4$ form a magnetic period. Under external $H$, the local linear spin susceptibility tensor $A_i^{\alpha l}$ of spin at atom $i$ ($i$=1-4) has only three non-zero components by excluding the Van-fleck paramagnetism, i.e., $A_i^{xx} = A_i^{yy} = \chi_{i\perp}, A_i^{zz} = \chi_{i\parallel}$. Such simplifying process is based on the Wyckoff symmetry $\mathbf{3}_z$ of Cr atoms. Based on the crystal symmetries, we found that all the $S_i$ have identical $\chi_{i\perp}=\chi_\perp$ and $\chi_{i\parallel}=\chi_\parallel$.

Then, we can determine the single spin tensor $P_{ii}^{\alpha\beta\gamma}$ by Cr moment at atom $i$ ($i$=1-4). Based on the global $\mathbf{3}_z$ symmetry in the crystal lattice, the symmetric matrix form of $P_{11}^{\alpha\beta\gamma}$ can be simplified as following:

$$P_{11}^{\alpha\beta\gamma} = \begin{pmatrix} P_{ii}^{xxx}, P_{ii}^{xxy}, P_{ii}^{xxz} & P_{ii}^{xxy} & -P_{ii}^{xxx} & 0 & P_{ii}^{xzx} & P_{i,i}^{xzy} & 0 \\ P_{ii}^{xxy} & -P_{ii}^{xxx} & 0 & -P_{ii}^{xxx} & -P_{ii}^{xxy} & P_{ii}^{xxz} & -P_{ii}^{xzy} & P_{ii}^{xzx} & 0 \\ P_{ii}^{xzx} & P_{ii}^{xzy} & 0 & & -P_{ii}^{xzy} & P_{ii}^{xzx} & 0 & 0 & 0 & P_{ii}^{zzz} \end{pmatrix} \quad (5)$$

The rest $P_{ii}^{\alpha\beta\gamma}$ ($i$=2,3,4) can be determined accordingly from symmetry transformation rule. As for the two-spin tensor $P_{ij}^{\alpha\beta\gamma}$ between adjacent atoms $i$ and $j$, there are five nearest exchange couplings $J_1$ to $J_5$ that contribute to the observed LME effects [20], as shown in Fig. 1(c), corresponding to the local two-spin tensors $P_{1,2}$, $P_{1,4''}$, $P_{1,2'}$, $P_{1,3'}$ and $P_{1,3}$ respectively. The related $P_{ij}^{\alpha\beta\gamma}$ were simplified by symmetry operations as well (see SM for details [16]).

Finally, the polarization $P$ at zero magnetic field $H=0$, and $\delta P$ that are induced by $\delta H$ along $x$ and $z$ directions of $Cr_2O_3$ are calculated by substituting all local ME tensors, the initial spin vectors $S_i$ and $A_i^{\alpha l}$ into Eq. (4) (see SM for details [16]). The first term of $P$ under zero $H$ is calculated to be exactly zero, consistent with the LME nature of $Cr_2O_3$. The second LME terms show finite response of polarization $\delta P$ with $\alpha_\perp = \delta P^x/\delta H^x$ and $\alpha_\parallel = \delta P^z/\delta H^z$:

$$\alpha_\perp \propto \left(2P_{11}^{xzx} - 2P_{12}^{xzx} + 2P_{13'}^{zxx} + 2P_{13'}^{xzx} + P_{12'}^{zxx} - P_{12'}^{xzx} + P_{14''}^{zxx} - P_{14''}^{xzx}\right)\chi_\perp S^z, \quad (6)$$

$$\alpha_\parallel \propto \left(P_{11}^{zzz} + P_{13'}^{zzz}\right)\chi_\parallel S^z \quad (7)$$

where $P_{11}^{zzz}$ and $P_{11}^{xzx}$ come from single-spin tensor, while the rest coefficients are originated from the two-spin tensors. One can see that, $\alpha_\parallel$ and $\alpha_\perp$ are proportional to $\chi_\parallel$ and $\chi_\perp$, respectively, which are fully consistent with previous theoretical studies and experimental observations. More importantly, both single spin tensor and two-spin tensor contribute to $\alpha_\parallel$ and $\alpha_\perp$, which is not anticipated in the previous studies [13,17-21] that neglect single-spin terms as: $p_{ii}^x = P_{ii}^{xzx}(\delta S_i^x S_i^z + S_i^x \delta S_i^z)$ and $p_{ii}^z = 2P_{ii}^{zzz}\delta S_i^z S_i^z$. From the matrix form of $p$-$d$ hybridization, both terms are allowed (see SM for details [16]). In terms of two-spin tensors, $P_{13'}^{zzz}$ is allowed in the exchange striction mechanism, while $(P_{12'}^{zxx} - P_{12'}^{xzx})$, $(P_{14''}^{zxx} - P_{14''}^{xzx})$ and $P_{12}^{xzx}$ are non-zero in the inverse DM mechanism. Note that, the $P_{13'}^{zxx} + P_{13'}^{xzx}$ is exactly zero in both two-spin mechanisms (see SM for details [16]), but is allowed in the antisymmetric exchange mechanism proposed by Xiang [8]. To conclude for $Cr_2O_3$, our approach suggests that $\alpha_\perp$ is derived by the *p-d* hybridization mechanism with single spin, and the inverse DM coupling mechanism from $J_1$, $J_2$ and $J_3$, possibly with small contribution from $J_4$ with other mechanisms. The $\alpha_\parallel$, on the other hand, is mainly controlled by the *p-d* hybridization and the exchange mechanisms from $J_4$. All the calculated results are summarized in Table II. the Our findings are consistent with previous studies but provides much better details down to atomic scale [18,21].

We then apply this approach to the multiferroic material with possible QME effects, the manganites *R*MnO$_3$ with orthorhombic crystal structure (*o-R*MnO$_3$), where *R* is rare-earth ion. *o-R*MnO$_3$ has the *Pbnm* space group, including symmetric operations of inversion symmetry {-1| 0,0,0}, two-fold rotation along *z*-axis {$2_{001}$ | 1/2,0,1/2} and two-fold rotation along *y*-axis {$2_{010}$ | 0,1/2,0}, as shown in FIG. 2. The neutron diffraction study and pyroelectric measurements confirm that below 40 K, *o-R*MnO$_3$ (*R*=Lu, Ho, Gd, Eu…) polycrystalline sample shows ferroelectricity with E-type AFM ordering of Mn spins [14], as shown in FIG. 2(a). Theoretical predictions have indicated that E-type AFM can break the space inversion symmetry and bring out huge ferroelectric polarization in *o-R*MnO$_3$ via exchange striction mechanism [14,22] which is partially confirmed in experiments [14,15]. Dominating quadratic ME effects were clearly observed in both poly and single crystal *o-R*MnO$_3$ systems in experiments [14,15]. However, the existence of ME effects in this system are not well explained on neither magnetic symmetry nor down to microscopic level.

We first consider whether the magnetic point of E-type AFM allows the ME effects. The magnetic point group of *o-R*MnO$_3$ with Mn E-type AFM ground state is **mm21'**, which is a polar group supporting ferroelectricity. Here, we do not consider the ordering of *R* ions. It belongs to magnetic "grey group" with an independent time reversal operation, which is unable to generate any LME effect but QME effect, consistent with the experimental observations [14,15]. The expected QME tensor $\beta^{ijk}$ can be written down in an abbreviated matrix form of:

$$\beta^{ijk} = \begin{pmatrix} 0 & 0 & 0 & 0 & Q_{15} & 0 \\ 0 & 0 & 0 & Q_{24} & 0 & 0 \\ Q_{31} & Q_{32} & Q_{33} & 0 & 0 & 0 \end{pmatrix}. \qquad (8)$$

It is symmetrical with $\beta^{ijk}=\beta^{ikj}$ and we follow the convention: $Q_{31}=\beta^{zxx}$, $Q_{32}=\beta^{zyy}$, $Q_{33}=\beta^{zzz}$, $Q_{24}=\beta^{yyz}$, $Q_{15}=\beta^{xxz}$.

To calculate the components $\beta^{ijk}$ from our theory, the starting AFM spin configuration at *H*=0 would be $S_{\text{up}}=(S^x,0,0)$ and $S_{\text{down}}=(-S^x,0,0)$. Under small external $\delta H$, the spin at atom *i* has local linear and quadratic spin susceptibility tensors $A_i^{\alpha l}$ and $B_i^{\alpha lm}$, respectively. No simplification can be made on those local susceptibility tensors. As for the single spin tensor $P_{ii}^{\alpha\beta\gamma}$ for Mn atom at site *i*, it will be exactly zero due to the onsite inversion symmetry. The measured polarization and QME in *o-R*MnO$_3$ are merely due to Mn-Mn interactions. All the spin pairs related to $P_{ij}^{\alpha\beta\gamma}$ in the magnetic unit cell can be deduced from symmetry operations (see SM for details [16]).

Similarly, following the procedure of Cr$_2$O$_3$, we can evaluate *P*(*H*=0) and $\delta P$ using Eq. 4 by inserting the two-spin tensors, the initial spin vectors $S_{\text{up}}$ and $S_{\text{down}}$, and $A_i^{\alpha l}$ and $B_i^{\alpha lm}$ (see SM for details [16]). The first term in *P*(*H*=0) reads out easily as: *P*∝-

$8(S^x)^2(0,0,P_{12}^{xxz})$, with $P_{12}^{xxz}$ originating from exchange striction between intralayer interactions, agreeing well with previous theoretical studies [14]. The second term in the LME of $\delta P$ appears to be exactly zero, which is consistent with the magnetic point group analysis. Finally, finite QME coefficients emerge from the third terms:

$$\beta^{zxx} \propto -8S^x\left[2P_{12}^{xxz}B_1^{xxx} + \left(P_{12}^{xyz} + P_{12}^{yxz}\right)B_1^{yxx} + (P_{12}^{zxz} - P_{12}^{xzz})B_1^{zxx}\right]$$

$$\beta^{zyy} \propto -8S^x\left[2P_{12}^{xxz}B_1^{xyy} + \left(P_{12}^{xyz} + P_{12}^{yxz}\right)B_1^{yyy} + (P_{12}^{zxz} - P_{12}^{xzz})B_1^{zyy}\right]$$

$$\beta^{zzz} \propto 16S^x\left[2P_{12}^{xxz}B_1^{xzz} + \left(P_{12}^{xyz} + P_{12}^{yxz}\right)B_1^{yzz} + (P_{12}^{zxz} - P_{12}^{xzz})B_1^{zzz}\right]$$

$$\beta^{xxz} \propto 16S^x\left[(P_{12}^{xyx} - P_{12}^{yxx})B_1^{yxz} - (P_{12}^{xzx} + P_{12}^{zxx})B_1^{zxz}\right]$$

$$\beta^{yyz} \propto 16S^x\left[(P_{12}^{xyy} - P_{12}^{yxy})B_1^{yyz} - (P_{12}^{xzy} + P_{12}^{zxy})B_1^{zxz}\right] \quad (9)$$

which is well consistent with magnetic point group argument. Note that, $A_i^{\alpha l}$ and interlayer interactions do not contribute to the QME in $o$-$R$MnO$_3$ directly. In terms of $B_1^{\alpha lm}$ for spin 1 with (-$S^x$,0,0), we could assume that under small $\delta H$ along the principle axis, the AFM spin configuration will change according to the FIG. 1(b) by simply changing their magnitude or rotating in $xz$ or $xy$ planes. As a result, for $H//x$, $B_1^{xxx}$, $B_1^{yxx}$, $B_1^{zxx} = 0$, for $H//y$ and $z$, $B_1^{yxz}$, $B_1^{zyy}$ and $B_1^{yzz} = 0$ due to constraint of pure spin rotating, $B_1^{yyy}$ and $B_1^{zzz} = 0$ due to ideal linear $\chi_\perp$, while $B_1^{yyz}$, $B_1^{zxz}$, $B_1^{xyy}$ and $B_1^{xzz}$ can be nonzero. It can be easily found that $B_1^{xyy} = (A_1^{yy})^2/2S^x$ and $B_1^{xzz} = (A_1^{zz})^2/2S^x$, respectively while $B_1^{yyz}$ and $B_1^{zxz}$ would be relatively much smaller. From above assumptions, $\beta^{zxx}$ must be zero. In order to find out the dominating mechanisms responsible for the QME observed in experiments, we need to find out the possible mechanisms for each $P_{ij}^{xxz}$ and the relative magnitude of $B_1^{\alpha lm}$ in AFM orders. In terms of two spin tensors, $P_{12}^{xxz}$ is allowed in the exchange striction mechanism while $(P_{12}^{xyy} - P_{12}^{yxy})$ is non-zero in inverse DM mechanism. Note that, the $(P_{12}^{xzx} + P_{12}^{zxx})$ and $(P_{12}^{xzy} + P_{12}^{zxy})$ are exactly zero in both mechanisms, but allowed in antisymmetric exchange mechanism proposed by Xiang [8]. In most cases, they are very small. Eventually, the dominating non-zero $\beta^{ijk}$ would be: $\beta^{zyy} \propto P_{12}^{xxz}(A_1^{yy})^2$, $\beta^{zzz} \propto P_{12}^{xxz}(A_1^{zz})^2$. $\beta^{zyy} \propto S^x(P_{12}^{xzx} + P_{12}^{zxx})B_1^{xxz}$ and $\beta^{yyz} \propto S^x[(P_{12}^{xyy} - P_{12}^{yxy})B_1^{yyz} - (P_{12}^{xzy} + P_{12}^{zxy})B_1^{zxz}]$ would be very small. All the calculated results are summarized in Table II.

We can compare our results with the experimental data. In the single crystal $o$-TbMnO$_3$ under pressure, it also has an E-type AFM ground state. The external $H//z$ can induce a huge, quadratic change of $P$ along $z$ above Tb ordering temperature, indicating a non-zero $\beta^{zzz}$. In the polycrystalline $o$-LuMnO$_3$ [14], we found that the change of $\delta P$ under 14 T is almost invariant below 20 K, as shown in Fig. 3, which is consistent with the almost constant $\chi_\perp$ in AFM orders. In conclusion, the QME of $o$-$R$MnO$_3$ mainly comes from exchange striction mechanism same as that of its original spin induced polarization and are proportional to the square of $\chi_\perp$.

In summary, we develop a general approach to explore the microscopic mechanism of spin-induced ME effects in single phase multiferroics and magnetoelectrics. Our findings are consistent with experiments and previous studies with much better details down to atomic scale.

**ACKNOWLEDGEMENTS**:

This work is supported by the Natural Science Foundation of China grant Nos. 11674384, 11974065. This work has been supported by Chongqing Research Program of Basic Research and Frontier Technology, China (Grant No. cstc2020jcyj-msxmX0263), Fundamental Research Funds for the Central Universities, China (2020CDJQY-A056, 2020CDJ-LHZZ-010, 2020CDJQY-Z006), Projects of President Foundation of Chongqing University, China(2019CDXZWL002).

**REFERENCE**

1. Y. Choi, H. Yi, S. Lee, Q. Huang, V. Kiryukhin and S. W. Cheong, Phys. Rev. Lett. **100**, 047601 (2008).

2. T. Kimura, Y. Seiko, H. Nakamura, T. Siegrist and A. Ramirez, Nature Mater. **7**, 291–4 (2008).

3. H. Murakawa, Y. Onose, S. Miyahara, N. Furukawa and Y. Tokura, Phys. Rev. Lett. **105**, 137202 (2010).

4. Y. S. Chai, S. H. Chun, J. Z. Cong, and K. H. Kim, Phys. Rev. B **98**, 104416 (2018).

5. H. J. Xiang, E. J. Kan, Y. Zhang, M.-H. Whangbo, and X. G. Gong, Phys. Rev. Lett. **107**, 157202 (2011).

6. T. A. Kaplan and S. D. Mahanti, Phys. Rev. B **83**, 174432 (2011).

7. S. Miyahara and N. Furukawa, Phys. Rev. B **93**, 014445 (2016).

8. J. S. Feng and H. J. Xiang, Phys. Rev. B **93**, 174416 (2016).

9. D. S. Shang, Y. S. Chai, Z. X. Cao, J. Lu, and Y. Sun, Chin. Phys. B **24**, 068402 (2015).

10. S. H. Chun *et al*., Phys. Rev. Lett. **108**, 177201 (2012).

11. G. Srinivasan, Annu. Rev. Mater. Res. **40** 153–78 (2010).

12. S. Dong, J. M. Liu, S. W. Cheong and Z.F. Ren, Adv. Phys. **64**, 519 (2015).

13. V. J. Folen, G. T. Rado, and E. W. Stalder, Phys. Rev. Lett. **6**, 607 (1961).

14. S. Ishiwata, Y. Shintaro, Y. Tokunaga, Y. Taguchi, T. Arima and Y. Tokura, Phys. Rev. B **81**, 100411 (2010).

15. T. Aoyama, K. Yamauchi, A. Iyama, S. Picozzi, K. Shimizu and T. Kimura, Nat. Commun. **5**, 4927 (2014).

16. See Supplemental Material for the details of symmetry analysis, and tensor and polarization calculations.

17. D. N. Astrov, Sov. Phys. JETP, **13,** 729-733 (1961).

18. M. Mostovoy, A. Scaramucci, N. A. Spaldin, and K. T. Delaney, Phys. Rev. Lett. **105**, 087202 (2010).

19. G. T. Rado, and V. J. Folen, J. Appl. Phys. **33**, 1126 (1962).

20. E. J. Samuelsen, M. T. Hutchings and G. Shirane, Physica **48**, 13-42 (1970).
21. R. Hornreich, S. Shtrikman, Phys. Rev. **161**, 506-512 (1967).
22. S. Picozzi, K. Yamauchi, B. Sanyal, I. A. Sergienko and E. Dagotto, Phys. Rev. Lett. **99**, 227201 (2007).

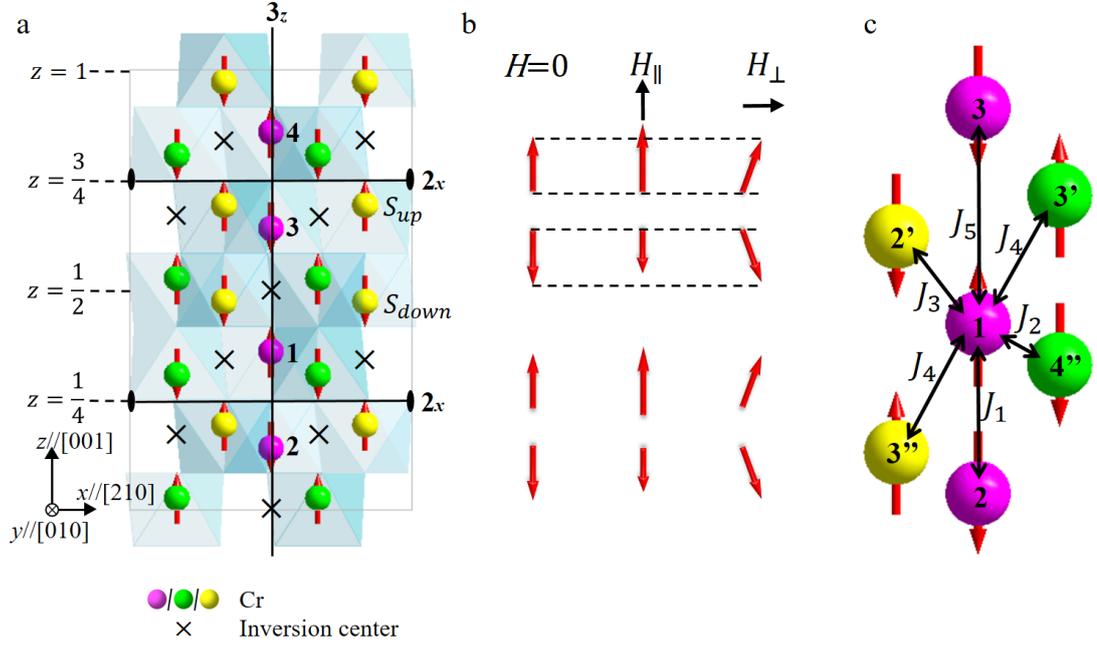

FIG. 1. (a) The crystal structure and AFM structure of $Cr_2O_3$. Space inversion centres where existing at the adjacent two Cr atoms and 3-fold axis along $z$ direction are indicated in the figure. (b) Considering the spins variations of four purple Cr atoms under different magnetic field applied, when ***H*** applied in AFM domain direction, the parallel spins or antiparallel spins tend to elongate or shorten, respectively. When the ***H*** perpendicular to AFM domain, all spins deflect to the same direction. (c) Spin-bonds($J_1$-$J_5$) between every Cr atom with its six nearest neighbours, of which $J_2$, $J_4$ and $J_3$ are all three bonds connected with three-fold symmetry in $z$ direction.

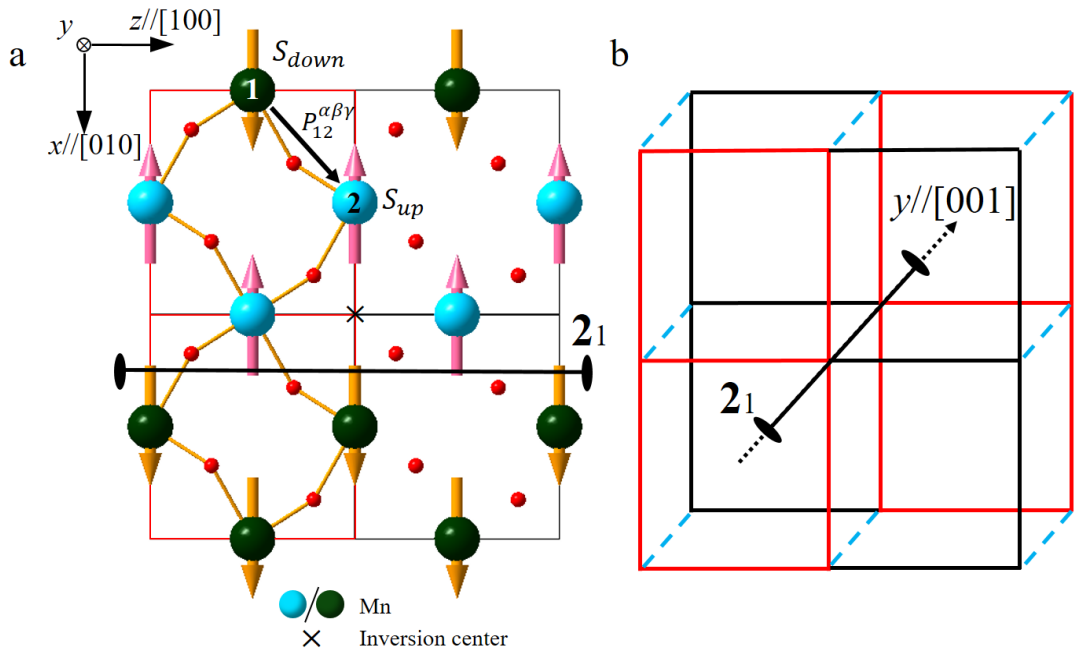

FIG. 2. (a) The configuration of Mn and oxygen atoms in *xz* plane in *o-R*MnO$_3$ lattice; (b) **2**$_1$ symmetry operation between two atom layers;

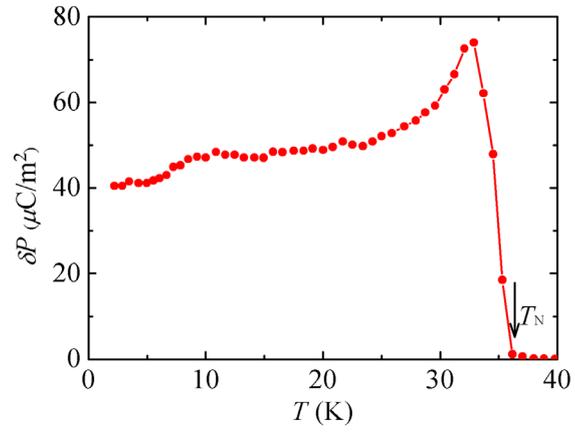

FIG. 3. The variation of polarization $\delta P = P(0\ \text{T}) - P(14\ \text{T})$ under 14 T in polycrystalline $o$-LuMnO3 [14].

Table I. Required local information to calculate macroscopic physical properties.

| | $P_{ii}^{\alpha\beta\gamma}$ or $P_{ij}^{\alpha\beta\gamma}$ | $S_i$ | $A_i^{\alpha l}$ | $B_i^{\alpha lm}$ |
|---|---|---|---|---|
| Spin-induced ferroelectricity | √ | √ | | |
| LME effect | √ | √ | √ | |
| QME effect | √ | √ | √ | √ |

Table II. The calculated nonzero single or two-spin tensors, LME coefficients, QME coefficients in $Cr_2O_3$ and $R$MnO$_3$ and the related microscopic mechanisms.

| | Exchange striction | Inverse DM interaction | p-d hybridization | Anisotropic symmetric exchange |
|---|---|---|---|---|
| $Cr_2O_3$ | $P_{13'}^{zzz}: \alpha_\parallel, J_4$ | $(P_{12'}^{zxx} - P_{12'}^{xzx}): \alpha_\perp, J_3$<br>$(P_{14''}^{zxx} - P_{14''}^{xzx}): \alpha_\perp, J_2$<br>$P_{12}^{xzx}: \alpha_\perp, J_1$ | $P_{11}^{xzx}: \alpha_\perp$<br>$P_{11}^{zzz}: \alpha_\parallel$ | $P_{13'}^{zxx} + P_{13'}^{xzx}: \alpha_\perp, J_4$ |
| $o$-$R$MnO$_3$ | $P_{12}^{xxz}: \beta^{zyy}, \beta^{zzz}$ | $(P_{12}^{xyy} - P_{12}^{yxy}): \beta^{yyz}$ | | $(P_{12}^{xzx} + P_{12}^{zxx}): \beta^{zyy}$<br>$(P_{12}^{xzy} + P_{12}^{zxy}): \beta^{yyz}$ |